\documentclass[preprint,prl,aps,amsmath,amssymb,color,superscriptaddress]{revtex4}
\usepackage{stmaryrd}
\usepackage{amsmath}
\usepackage{amssymb}
\usepackage{graphicx}
\usepackage{dcolumn}
\usepackage{bm}
\usepackage{multirow}
\usepackage{tabularx}
\usepackage{epsfig}
\usepackage{url}

\begin{document}
\begin{titlepage}
\title{Optical signature for distinguishing between Mott-Hubbard, intermediate and charge-transfer insulators}
\author{Xiangtian Bu}
\affiliation {Key Lab of advanced optoelectronic quantum architecture and measurement (MOE), and Advanced Research Institute of Multidisciplinary Science, Beijing Institute of Technology, Beijing 100081, China}
\author{Yuanchang Li}
\email{yuancli@bit.edu.cn}
\affiliation {Key Lab of advanced optoelectronic quantum architecture and measurement (MOE), and Advanced Research Institute of Multidisciplinary Science, Beijing Institute of Technology, Beijing 100081, China}

\date{\today}

\begin{abstract}
Determining the nature of band gaps in transition-metal compounds remains challenging. We present a first-principles study on electronic and optical properties of CoO using hybrid functional pseudopotentials. We show that optical absorption spectrum can provide a clear fingerprint to distinguish between Mott-Hubbard, intermediate and charge-transfer insulators. This discrimination is reflected by the qualitative difference in peak satellites due to unique interplay between $d$-$d$ and $p$-$d$ excitations, thus allowing identification from experimental data alone, unlike the existing methods that require additional theoretical interpretation. We conclude that the CoO is an intermediate, rather than a Mott-Hubbard insulator as is initially believed.
\end{abstract}

\maketitle
\draft
\vspace{2mm}
\end{titlepage}

Transition-metal compounds form a rich and fruitful field of research, not only for its scientific significance in high-$T_c$ superconductivity, colossal magnetoresistance, phase transition and multiferroicity\cite{Khomskii,Imada,Lee,MARIANETTI2004}, but also for its commercial applications in batteries and catalysts\cite{Liu2018,Ozawa1994,Westerhaus2013,Iravani2020}. Explaining these phenomena and improving device performance require a quantitative understanding of their fundamental electronic structures, particularly the nature of the gap, which is associated with the competition between on-site Coulomb repulsion $U$ of $d$-orbitals and charge-transfer energy $\Delta$ from anion to cation. According to the Zaanen-Sawatzky-Allen diagram\cite{Zaanen1985}, if $U < \Delta$, the gap is of Mott-Hubbard type; If $U > \Delta$, the gap is of charge-transfer type; If $U\approx \Delta$, the gap is of intermediate type. As schematically illustrated in Fig. 1, different gaps correspond to completely different lowest charge excitations, such as $d$-$d$ and $p$-$d$ excitations for Mott-Hubbard and charge-transfer insulators, respectively. This distinction is particularly important for high-$T_c$ superconductivity in the case of hole doping\cite{Lee,Zhang,PNAS}, because it determines whether the doped holes predominantly sit in the metal-$d$ or anion-$p$ state.

However, unambiguously determining the nature of the gap remains challenging\cite{Velasco2011}. There has been no experimental way to straightforwardly derive $U$ and $\Delta$. Although they can be derived from electronic structure calculations, the accuracy is not yet satisfactory when dealing with partially filled $d$-electron systems\cite{TanJCP2019}. Popular methods such as density functional theory (DFT) often fail even to open energy gaps. To remedy the problem, several post-DFT methods have been developed, which include tunable parameters that need to be fitted experimentally. This, in turn, requires quantitative agreement between experiment and theory, which is almost impossible. Different manners whereby the parameters are fitted further increase the uncertainty. As a result, different interpretations of experimental data from scanning tunneling microscopy/photon-electron spectroscopy give conflicting results in the literature\cite{Brookes1989,Shen1990,Nakata,LiuLW,LiuMK,LiuZY}. A qualitatively discriminating signature, especially the kind that can be directly observed from experiments without theoretical assistance, naturally becomes a key issue to be addressed in theoretical studies, which will be the focus here.

The difference in the lowest excitation leads us to the optical scheme (see Fig. 1). According to the Laporte selection rule\cite{Laporte}, the $d$-$d$ transition is optically-forbidden while the $p$-$d$ transition is optically-allowed for a centrosymmetric complex. Although, in solids, this selection rule is partially relaxed as a result of orbital hybridization, such differences still substantially affect the optical properties\cite{Jiang2019}, probably leaving fingerprints in the absorption spectrum or the exciton structure. Herein, we will take cobalt monoxide (CoO) as an example for illustration. CoO is theoretically classified as a Mott-Hubbard insulator\cite{Brandow}, but an increasing number of experiments point to a pivotal role of charge transfer\cite{Shen1990,Lee1991,Elp1991,Magnuson2002}. At present, to determine the nature of the CoO gap, people first need its experimental value to calibrate the tunable parameter in the calculations and then derive $U$ and $\Delta$ from the corresponding band structure. However, two facts limit the usefulness of this scheme: i) The reported gaps of CoO span a wide range from 2.5 to 6.0 eV\cite{Elp1991, Kang2007, Pratt1959, Shen1990, Gvishi1972}. ii) Some of the adopted calculations contain more than one parameter that cannot be determined using only one gap value\cite{Fuchs2009,Lanata2019}. Therefore, the nature of the CoO gap remains elusive to today. On the other hand, although optical absorption of CoO has been measured more than 60 years ago\cite{Pratt1959}, its explicit interpretation is still absent.

%fig01
\begin{figure}[htbp]
\includegraphics[width=0.7\columnwidth]{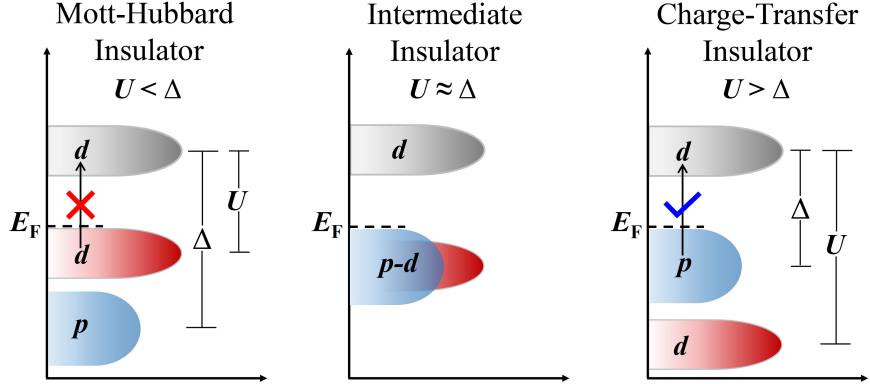}
\caption{\label{fig:fig1} Schematics of Mott-Hubbard, intermediate and charge-transfer insulators, as well as the different lowest excitations. The $d$-$d$ transition is optically-forbidden while the $p$-$d$ transition is optically-allowed.}
\end{figure}

In this work, we seek to reveal qualitative differences in the optical spectra of Mott-Hubbard, intermediate and charge-transfer insulators. To this end, we employ the recently developed Co hybrid functional pseudopotentials which has been shown to open for CoO a gap at the DFT level\cite{TanJCP2019}. With the increase of exact exchange in generating hybrid functional pseudopotentials, we find that the resulting CoO band structure shows a continuous transition from Mott-Hubbard to intermediate and then to charge-transfer insulating phase. This allows us to directly compare the optical absorption features of the three different insulators, which show significant differences in terms of peak satellites. In particular, the absorption spectrum of the Mott-Hubard phase contains only a two-peak skeleton while a third satellite peak appears between the two peaks in the case of the intermediate phase. This satellite peak in the charge-transfer phase further evolves into a plateau. Thus, a simple light absorption experiment will fully reveal the fingerprints of Mott-Hubard and charge-transfer physics in transition-metal compounds. Comparison with available experimental data suggests that CoO is an intermediate rather than a Mott-Hubbard insulator.

All DFT calculations were carried out using the Quantum Espresso package\cite{Giannozzi2017} with the Perdew-Burke-Ernzerhof (PBE) \cite{Perdew1996} exchange-correlation functional. Hybrid functional pseudopotentials\cite{Jing2018} generated from the OPIUM\cite{opium} code were used for Co while norm-conserving Vanderbilt pseudopotentials \cite{Hamann,SCHLIPF2015} were used for O. Note that here the $\alpha$ which specifies the exact exchange weight of the exchange-correlation functional used in hybrid pseudopotential construction is taken as an adjustable parameter\cite{TanJCP2019,Jing2018}. An energy cutoff of 60 Ry was set after a convergence test. A $k$-grid of $12 \times 12 \times 12$  was employed to sample the Brillouin zone throughout this work. The optical adsorption spectrum was obtained by solving the Bethe-Salpeter equation (BSE) \cite{Rohlfing2000} as implemented in the YAMBO code \cite{marini2009yambo,Sangalli2019}. A total of 300 empty bands were used to calculate the dielectric function matrix, with 45 and 12 Ry cutoff respectively for exchange and correlation parts. Top two valence bands and bottom two conduction bands were considered to construct the BSE Hamiltonian. Test calculations involving more bands yield the same conclusion. Here studied CoO has the rock salt structure with antiferromagnetic alignment along [111] direction\cite{Jauch2001}. The cubic lattice parameter is fixed at the experimental value 4.26 \AA.

We first calculate the electronic structures of CoO using different Co hybrid functional pseudopotentials with the $\alpha$ ranging from 0.001 to 1, and selectively plot the results in Fig. 2(a) for $\alpha$ = 0.001, 0.25, 0.50 and 0.75\cite{note}. These bands look very similar to each other, except for the parabolic O-2$p$ unoccupied band (highlighted in red) which continues to move up as the $\alpha$ increases. It constitutes the conduction band bottom at $\alpha$ = 0.001, analogous to previous studies\cite{Fuchs2009,Engel}. Whilst, the parabolic O-2$p$ band crosses the lowest Co 3$d$-band at $\alpha$ = 0.25 and all three Co 3$d$-bands at $\alpha$ = 0.75.

%fig02
\begin{figure*}[htbp]
\begin{center}
\includegraphics[width=0.95\columnwidth]{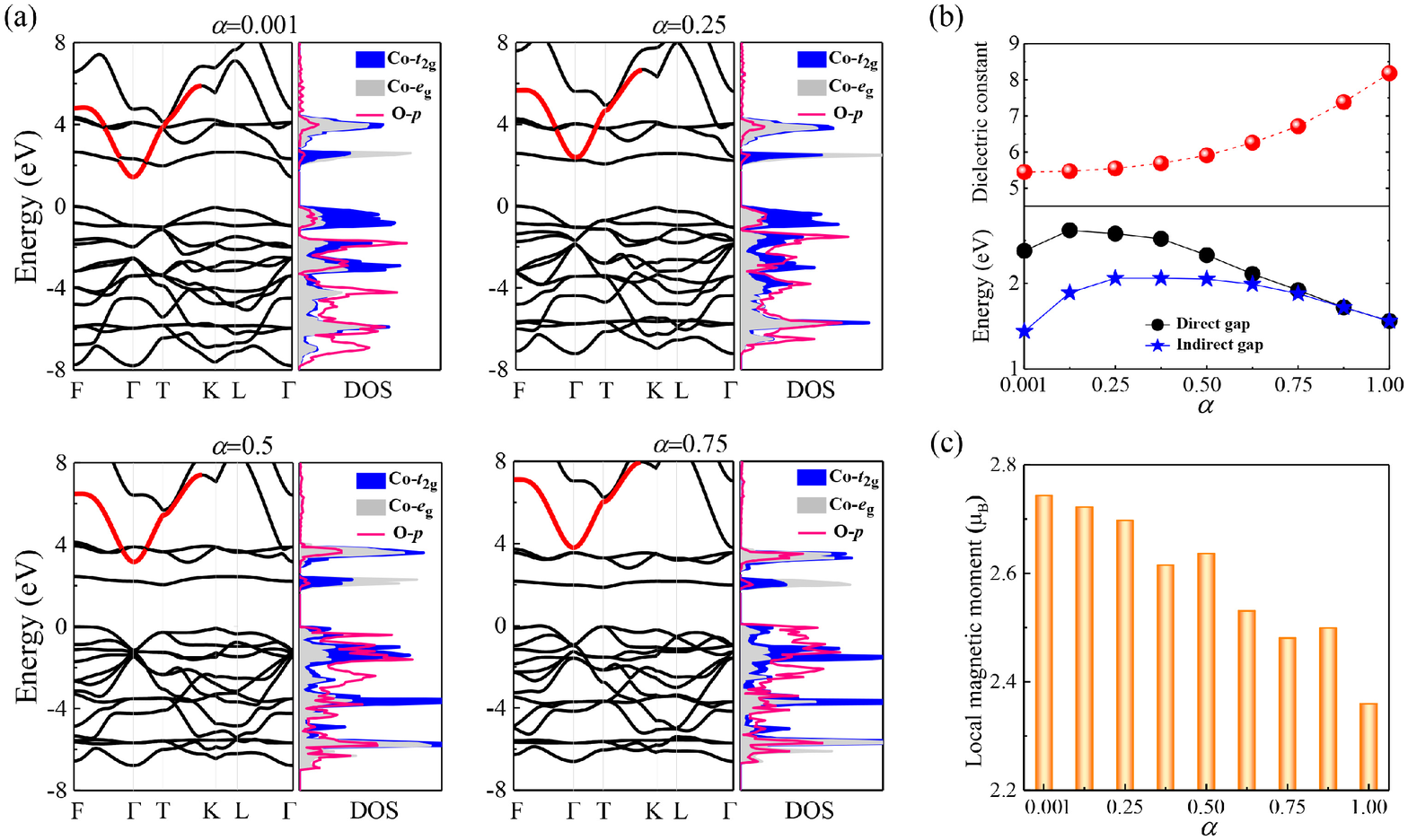}
\caption{\label{fig:fig2}(Color online) (a) Band structures (Left) and projected density of states (Right) of CoO at typical $\alpha$. Here only one spin channel is shown. For the antiferromagnetic case, the other spin is fully degenerate. The valence band maximum is set to energy zero. (b) Dielectric constant, direct- and indirect-gap, and (c) local magnetic moment of Co as a function of $\alpha$.}
\end{center}
\end{figure*}

All bands are semiconducting with a fundamental gap $>$ 1.4 eV. It is noteworthy that only a very small amount of exact exchange, e.g., $\alpha$ = 0.001, is sufficient to yield a gap. This is very interesting because even when applied to antiferromagnetic configuration, the PBE fails to produce a gap in CoO, unlike the case in MnO and NiO\cite{Terakura1984,Wei1994,Gopal2017,Fabien2006,LiuJJ2019}. While it is still debated whether orbital-dependent potentials are necessary\cite{Terakura1984,Engel}, it is recognized that the gap opening in CoO originates from the population imbalance among the Co minority $t_{2 \rm g}$ orbitals, as we obtain here. The difference between our calculation and the standard DFT is all in the way the Co core-electrons are treated. While the physics behind the gap opening is attractive to the DFT community, it belongs to another work outside the scope of the current study and will be discussed in the future.

As $\alpha$ increases, several interesting aspects are found. Firstly, there is a counter-intuitive correlation between the tendency of the fundamental gap and the dielectric constant (electronic contribution) to vary with $\alpha$. Generally speaking, the gap affects the system screening: the larger the gap, the weaker the screening and therefore the smaller the dielectric constant. This should have led to a synchronous variation relationship between the fundamental gap and the dielectric constant. However, this is not the case here. As can be seen in Fig. 2(b), either the minimum indirect gap or the direct gap shows a quadratic variation of increasing and then decreasing. Around $\alpha$ = 0.75, the gap undergoes an indirect-to-direct transition. The largest direct (indirect) gap reaches 2.62 (2.06) eV, very close to the photoemission result of 2.5 $\pm$ 0.3 eV\cite{Elp1991}. In contrast, the dielectric constant increases monotonically from 5.45 to 8.18, always larger than the experimental value of 5.35\cite{Powell1970}, as a result of the gap underestimation. This counter-intuitive gap-dielectric constant correspondence suggests that the electronic structure changes substantially during the $\alpha$ increase.

Secondly, the Co local magnetic moment decreases [see Fig. 2(c)]. Under the local crystal field with $O_h$ symmetry, the 3$d$-orbitals of Co split into a lower-lying $t_{2 \rm g}$ triplet and a higher-lying $e_{\rm g}$ doublet. In case of a Mott-Hubbard insulator, only the $d$-electron occupation is considered. Seven Co $d$-electrons fully occupy five states of the majority spin channel and the remaining two electrons partially fill the minority spin $t_{2 \rm g}$ triplet, giving rise to a local moment of 3 $\mu_{\rm B}$. In this sense, the 3 $\mu_{\rm B}$ local moment can be regarded as a sign of the Mott-Hubbard gap. Our calculations show that the moment decreases from 2.74 $\mu_{\rm B}$ at $\alpha$ = 0.001 to 2.36 $\mu_{\rm B}$ at $\alpha$ = 1. This decrease comes from the hybridization between the Co-3$d$ and the O-2$p$ states, and the hybridization is enhanced with increasing $\alpha$. Obviously, the charge-transfer role is becoming increasingly important. Note that the moments obtained here are comparable to other calculations\cite{LiuJJ2019}, but much smaller than the experimental 3.35$\sim$3.98 $\mu_{\rm B}$ \cite{Jauch2001,Khan1970,Ronzaud1978}, since the orbital contribution is neglected\cite{Fabien2006}.

Thirdly, the evolution of the projected density of states as $\alpha$ increases visualizes the enhancement of $p$-$d$ hybridization [see right panels in Fig. 2(a)], manifested by the increasing contribution of O-2$p$ to the valence band top. Its contribution increases from $\sim$10\% at $\alpha$ = 0.001 to 24\% at $\alpha$ = 0.25 and reaches $\sim$70\% at $\alpha$ = 0.75. At $\alpha$ = 0.5, the contributions of Co-3$d$ and O-2$p$ are about the same. On the other hand, the bottom conduction band is always dominated by the Co 3$d$-states. Referring to Fig. 1, the band undergoes a continuous transition from Mott-Hubbard-type to intermediate-type to charge-transfer-type as $\alpha$ increases. This is interesting because it provides a basis for exploring the optical properties related to the nature of the gap.

Next, we solve the BSE for the imaginary part of the dielectric function using the bands at different $\alpha$ ranging from 0.001 to 1 as inputs, and typical results are presented in Fig. 3(a). Experimental result extracted from Ref. \onlinecite{Pratt1959} is also added for comparison. Note that here we focus on the structural differences in the spectra given by different $\alpha$, so for clarity we uniformly adjust the first absorption peak (labeled as $p_1$) to energy zero (Its absolute energy corresponds to the optical gap as summarized in Fig. 3(b)). This is physically reasonable, since the gap underestimation can usually be handled by applying scissors during the solution of the BSE, which results in a rigid translation of the optical absorption spectrum along the energy axis without changing the spectral shape.

%fig03
\begin{figure}[htbp]
\includegraphics[width=0.75\columnwidth]{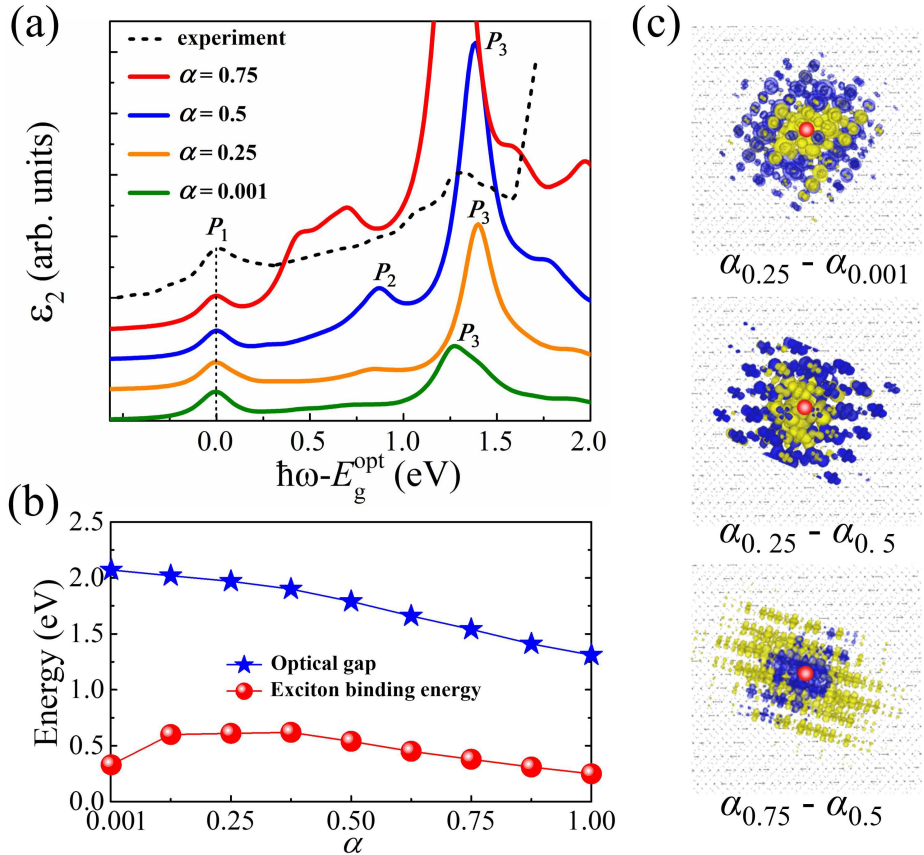}
\caption{\label{fig:fig3} (Color online) (a) Imaginary part of the dielectric function by solving the BSE at typical $\alpha$ (colored solid lines), with the experimental adsorption spectrum extracted from Ref. \onlinecite{Pratt1959} (black dash line) for comparison. The first peak $p_1$, whose absolute energy corresponds to the optical gap ($E_{\rm g}^{\rm opt}$), is aligned uniformly to the energy zero. (b) Optical gap and exciton binding energy as a function of $\alpha$. (c) Plots of the change of the real-space exciton wave-function under different $\alpha$. For instance, $\alpha_{0.25}-\alpha_{0.001}$ represents the difference between the exciton wave-functions at $\alpha$ = 0.25 and $\alpha$ = 0.001 with the holes placed at the same central position (Red balls). The electron accumulation and depletion regions are depicted in yellow and blue, respectively}
\end{figure}

Each spectrum contains a basic skeleton consisting of two peaks (labeled as $p_1$ and $p_3$). The intensity of $p_1$ does not change much as $\alpha$ increases, but the intensity of $p_3$ increases rapidly. Meanwhile, some satellite signals appear on the shoulder of $p_3$ and become more and more prominent as $\alpha$ increases. All spectra can be divided into three categories according to the different characteristics of such peak satellites: 1) When $\alpha \leq$ 0.25, the satellite is almost invisible; 2) When $\alpha\in$ (0.25, 0.75), a new adsorption peak (labeled as $p_2$) appears; 3) When $\alpha\geq$ 0.75, the peak of $p_2$ develops into a plateau connecting $p_1$ and $p_3$.

By comparing with the electronic structures shown in Fig. 2(a), we find that the change in the optical spectral features is correlated with the transition in the gap nature . When $\alpha\leq$ 0.25, the band is of Mott-Hubbard type, which corresponds to the two-peak structure of the optical spectrum. When $\alpha\in$ (0.25, 0.75), the band is of intermediate type, which corresponds to the three-peak structure of the optical spectrum. When $\alpha\geq$ 0.75, the band is of charge-transfer type, which corresponds to the structure of a plateau connecting two peaks in the optical spectrum. We observe a three-peak feature in the experimental spectrum of CoO. From above criterion we infer that CoO is an intermediate insulator. Nevertheless, we should note that quantitatively, we did not reproduce the experiment, including the energies of the characteristic absorption peaks and the separation between them.

Although the spectra here are calculated using CoO as an example, the differences in characteristic structures on the spectrum essentially stems from different excitations across different types of gaps (see Fig. 1). In other words, such differences depend on the properties of the band-edge states rather than on the elements. Therefore, a simple light absorption measurement would reveal these fingerprints and thus determine whether a transition-metal compound is a Mott-Hubbard, an intermediate or a charge-transfer insulator.

Different $d$-$d$ and $p$-$d$ transitions may also lead to different excitons, serving to distinguish different gap behaviors. Indeed, $p_1$ in Fig. 3(a) corresponds to an exciton inside the fundamental gap. Its energy defines the optical gap while its difference from the fundamental gap defines the exciton binding energy. Figure 3(b) summarizes the situation of both at different $\alpha$. As $\alpha$ increases, the optical gap decreases monotonically from 2.1 eV to 1.3 eV. This trend is consistent with a monotonic increase in the dielectric constant [see Fig. 2(b)], in contrast to the non-monotonic dependence of the fundamental gap. The exciton binding energy shows a complex variation, first increasing and then decreasing.

Generally speaking, the larger the exciton size, the more it shows an electron aggregation towards the outside away from the hole. By contrast, the smaller the exciton size, the more it shows an electron aggregation towards the hole in the center. In this sense, we fix the holes at the same central position and plot the change of the exciton wave-function for two different $\alpha$ in Fig. 3(c), whereby the exciton sizes are compared visually. One can see that as compared to the cases of $\alpha$ = 0.001 and $\alpha$ = 0.5, the exciton of $\alpha$ = 0.25 shows more electron aggregation toward the center, so it is smaller in size. This is consistent with its larger binding energy [see Fig. 3(b)]. When $\alpha\geq$ 0.5, the exciton binding energy continues to decrease and the corresponding exciton radius becomes larger and larger, as in the case of $\alpha$ = 0.75. Despite these differences, it seems no direct correspondence between the exciton feature and the nature of gap.

In summary, the electronic structure and optical properties of CoO has been reexamined using hybrid-functional pseudopotentials. The results show that CoO is an intermediate insulator rather than a Mott-Hubbard insulator as is initially believed. This conclusion is drawn based on our new knowledge that Mott-Hubbard, intermediate insulators and charge-transfer insulators have different optical spectral characteristics. In contrast to traditionally used methods that combine experimental gap as calibration with theoretical calculations, our new scheme allows direct discrimination from experimental data without relying on calculations, thus eliminating uncertainties caused by differences in calculation methods. In addition, the introduction of the exact exchange to core-electrons opens up the CoO gap within the PBE, which points to some as yet unrevealed role of core-electrons in the DFT gap problem. Our work provides new insights to understand the complex correlation effects and charge-transfer interactions in transition-metal compounds, especially for the interesting physical properties of cobalt oxides.

This work was supported by the Ministry of Science and Technology of China (Grant No. 2020YFA0308800) and the National Natural Science Foundation of China (Grant No. 12074034).


\begin{thebibliography}{90}
\bibitem{note} Actually, we performed calculations with $\alpha$ = 0.001, 0.125, 0.25, 0.375, 0.50, 0.625, 0.75, 0.875 and 1. The results show that 0.25 and 0.75 can serve as two boundary points between the different characteristic phases, either for the one-electron band structures or the optical spectra.
\end{thebibliography}
\end{document}